\documentclass{iauc}
\usepackage{graphicx}
\pubyear{2005}
\volume{199}
\pagerange{1--}
\jname{Probing Galaxies through Quasar Absorption Lines}
\editors{P. R. Williams, C. Shu, and B. M\'{e}nard, eds.}

\newcommand{\blHI}{\hbox{{\rm H}\kern 0.1em{\sc i}}}
\newcommand{\blHII}{\hbox{{\rm H}\kern 0.1em{\sc ii}}}
\newcommand{\blLya}{\hbox{{\rm Ly}\kern 0.1em$\alpha$}}
\newcommand{\blLyb}{\hbox{{\rm Ly}\kern 0.1em$\beta$}}
\newcommand{\blMgI}{\hbox{{\rm Mg}\kern 0.1em{\sc i}}}
\newcommand{\blMgII}{\hbox{{\rm Mg}\kern 0.1em{\sc ii}}}
\newcommand{\blFeII}{\hbox{{\rm Fe}\kern 0.1em{\sc ii}}}
\newcommand{\blCII}{\hbox{{\rm C}\kern 0.1em{\sc ii}}}
\newcommand{\blCIII}{\hbox{{\rm C}\kern 0.1em{\sc iii}}}
\newcommand{\blCIV}{\hbox{{\rm C}\kern 0.1em{\sc iv}}}
\newcommand{\blNV}{\hbox{{\rm N}\kern 0.1em{\sc v}}}
\newcommand{\blOVI}{\hbox{{\rm O}\kern 0.1em{\sc vi}}}
\newcommand{\blSiIV}{{\rm Si}\kern 0.1em{\sc iv}}
\newcommand{\blSiII}{\hbox{{\rm Si}\kern 0.1em{\sc ii}}}
\newcommand{\blNII}{\hbox{{\rm N}\kern 0.1em{\sc ii}}}

\title[Search For Organic Molecules] 
{A Search For Organic Molecules in Intermediate Redshift DLAs}

\author[Lawton {\etal}]   
{Brandon Lawton$^1$,
Brian York$^2$,
Sara L. Ellison$^2$, \\
Christopher W. Churchill$^1$
Rachel A. Johnson$^3$, 
\break \and Theodore P. Snow$^4$}

\affiliation{$^1$Department of Astronomy, New Mexico State University, \break 
Las Cruces, NM 88003, USA \break 
email: blawton@nmsu.edu, cwc@nmsu.edu \\[\affilskip]
$^2$Department of Physics and Astronomy, University of Victoria, \break 
Victoria, BC, V8W 3P6, Canada \break 
email: briany@uvic.ca, sarae@uvic.ca \\[\affilskip]
$^3$ Department of Astrophysics, Oxford University, \break
Oxford, OX1 3RH, UK \break
email: raj@astro.ox.ac.uk \\[\affilskip]
$^4$Center for Astrophysics and Space Astronomy, University of Colorado, \break
Boulder, CO 80309, USA \break
email: Theodore.Snow@Colorado.EDU}

\begin{document}

\maketitle

\begin{abstract}

There has been a renewed interest in searching for diffuse
interstellar bands (DIBs) due to their probable connection to organic
molecules and, thus, their possible link to life in the Universe.  Our
group is undertaking an extensive search for DIBs in DLAs via QSO
absorption-line systems.  Six of our DLA targets are presented here.
Our equivalent width (EW) limits for the {$\lambda$}5780 DIB line
strongly suggests that DIB abundance is below the Milky Way expected
value or that metallicity plays a large role in DIB strengths.

\keywords{quasars: absorption lines; (galaxies:) absorption lines;
galaxies: ISM, evolution; astrobiology; astrochemistry}

\end{abstract}

\firstsection 

\section{Introduction}

Diffuse interstellar bands (DIBs) are absorption features seen
abundantly in the Galaxy towards highly reddened stars, implying that
DIBs exist in dusty environments.  The current belief is that the
majority of the several hundred DIB lines discovered so far are due to
organic molecules such as polycyclic aromatic hydrocarbons (PAHs).
Due to the possible connection to the building blocks of life, DIBs
have experienced an enhanced level of interest in the last several
years.  Because of the difficulty in detecting their relatively weak
spectral features, extragalactic searches for DIBs using QSO
absorption-lines are more rare; however, there is one detection of the
{$\lambda$}4428 DIB line in a DLA at $z=0.5$ toward AO 0235+164
(\cite[Junkkarinen {\etal} 2004]{blref:Intro5}).

It would be desirable to put DIBs on the same statistical grounds as
metal line systems typically seen in QSO absorption-line studies.  In
our search for extragalactic sources of DIBs, we are investigating the
feasibility of using the redshift path density ($dN/dz$) to constrain
the evolution of organic molecules through cosmic time and of using
DIBs as a proxy for finding DLAs with $z<1$.  In this article we
examine six DLAs for the five strongest DIB lines ({$\lambda$}4428,
{$\lambda$}5780, {$\lambda$}5797, {$\lambda$}6284, and
{$\lambda$}6613) and place upper-limits on their strengths relative to
Galactic DIBs.

\begin{table}[h]
\begin{center}
\caption{Equivalent Width Limits of DIBs in DLAs}
\label{blawtontab:EW}
\begin{tabular}{cccllllll}\hline
    &           &          &         \multicolumn{6}{r}{\underbar{DIB $3\sigma$ Rest--Frame EW Limits, m{\AA}}} \\
  QSO & z$_{abs}$ & log $N({\blHI})$ & Facility &   {$\lambda$}4428 &    {$\lambda$}5780 & {$\lambda$}5797 & {$\lambda$}6284 & {$\lambda$}6613 \\\hline
0738+313 & 0.091 & 21.2                    & \bf{APO/DIS}           &     \bf{$<$37}     & \bf{$<$22$^a$} &  \bf{$<$21} &  \bf{$<$28} & \bf{$<$28}    \\
         &       &                         & MW Predicted           &                197 &            205 &          47 &         218 &         82    \\
         &       &                         & Z=$-$1 Scaled          &                 20 &             20 &           5 &          22 &          8    \\  

0738+313 & 0.221 & 20.9                    & \bf{APO/DIS}           &         \bf{$<$51} &     \bf{$<$22} &  \bf{$<$22} &  \bf{$<$30} &   \bf{...}    \\
         &       &                         & MW Predicted           &                 82 &             85 &          19 &          91 &               \\
         &       &                         & Z=$-$1 Scaled          &                  8 &              9 &           2 &           9 &               \\

0827+243 & 0.518 & 20.3                    & \bf{Keck/HIRES}        & \bf{$<$390$^{b}$}  &     \bf{$<$34} &  \bf{$<$34} &    \bf{...} &   \bf{...}    \\
         &       &                         & MW Predicted           &                 13 &             13 &           3 &             &               \\
         &       &                         & Z=$-$1 Scaled          &                  1 &              1 &        $<$1 &             &               \\ 

0952+179 & 0.239 & 21.0                    & \bf{WHT/ISIS}         &           \bf{...} &    \bf{$<$146} & \bf{$<$129} & \bf{$<$211} &   \bf{...}    \\
         &       &                         & MW Predicted           &                    &            324 &          74 &         346 &               \\
         &       &                         & Z=$-$1 Scaled          &                    &             35 &           7 &          34 &               \\ 

1127-145 & 0.313 & 21.7                    & \bf{Gem/GMOS}          &           \bf{...} &    \bf{$<$569} & \bf{$<$759} & \bf{$<$248} & \bf{$<$427}   \\
         &       &                         & \bf{VLT/UVES}          &         \bf{$<$63} &      \bf{ ...} &    \bf{...} &    \bf{...} &   \bf{ ...}   \\
         &       &                         & MW Predicted           &              1,053 &          1,093 &         249 &       1,167 &        436    \\
         &       &                         & Z=$-$1 Scaled          &                103 &            107 &          24 &         114 &         43    \\
         
1229-020 & 0.395 & 20.7                    & \bf{WHT/ISIS}         &           \bf{...} &    \bf{$<$262} & \bf{$<$185} & \bf{$<$475} &   \bf{...}    \\
         &       &                         & MW Predicted           &                    &             53 &          12 &          57 &               \\
         &       &                         & Z=$-$0.47 Scaled$^{c}$ &                    &             42 &           4 &          19 &               \\\hline
\multicolumn{9}{l}{$^{a}$ Sky Line - not included in limit} \\
\multicolumn{9}{l}{$^{b}$ Large Blend - unblended limit is 7 m{\AA}} \\
\multicolumn{9}{l}{$^{c}$ Metallicity from Boiss\'{e} {\etal} 1998} \\
\end{tabular}
\end{center}
\end{table}

\section{Results and Discussion}

The results for our observations and predictions are presented in
Table 1; the observations are in bold.  To arrive at our ``MW
Predicted'' limits for DIB strengths we use a known correlation where
the {$\lambda$}5780 equivalent width (EW) is proportional to the
neutral hydrogen column density, $N({\blHI})$, in Milky Way clouds
(\cite[Herbig 1993]{blref:Corr1}; \cite[York \etal, in prep]
{blref:Corr2}).  We use this relation to scale the Milky Way values to
the observed $N({\blHI})$ of the DLAs.  Our results are plotted in the
left panel of Figure 1 for the {$\lambda$}5780 DIB line.  Presented
are observed limits on the DLA DIB strength versus expected Milky Way
DIB strength for a cloud with the DLA $N({\blHI})$.  Those points
below the 1:1 correlation line correspond to DLAs that are deficient
in the {$\lambda$}5780 DIB based purely on Milky Way expectations.
Four of our six DLAs have {$\lambda$}5780 DIB strengths at least 0.5
dex below Milky Way strengths.  Points above the 1:1 correlation line
are unconstrained and additional data are required to obtain
meaningful limits on the DIB strengths.  For our entries in Table 1,
we applied a similar scaling to the other DIB lines using their
relative strengths in the Milky Way (\cite[Jenniskens \& Desert
1994]{blref:Corr2}).  The {$\lambda$}4428 DIB line is scaled by an
additional factor of 25{\%} based on the DLA at $z=0.5$ towards the
QSO AO 0235+164 (\cite[Junkkarinen {\etal} 2004]{blref:Intro5}).

\begin{figure}[t]
\includegraphics[width=5.3in,angle=0]{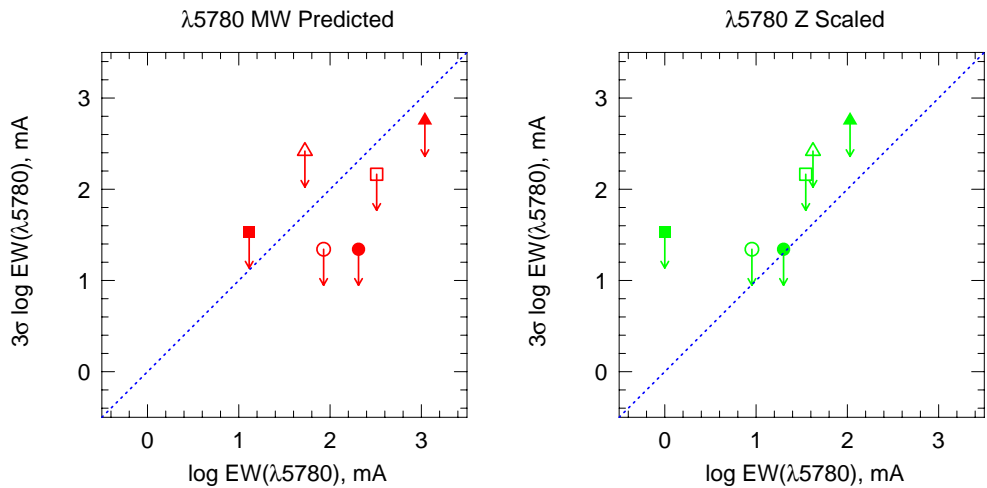}
\caption{--- (left) Measured 5780 DIB strength (upper limits)
vs. ``MW Predicted'' 5780 DIB strength.  --- (right) Same as left
except for ``Z-Scaled.''  Filled circle is 0738+313 for z=0.091.
Open circle is 0738+313 for z=0.221.  Filled square is 0827+243.  Open
square is 0952+197.  Filled triangle is 1127-145.  Open triangle is
1229-020.}
\label{blawtonfig:spectra}
\end{figure}

Including the DLA metallicity, we estimate the DIB
strengths by
\begin{equation}
\log EW = \log EW_{DIB} + \bigg( \log N({\blHI}) - 20.3 \bigg) + [\rm{Zn/H}] \mbox{  m{\AA}},
\end{equation}
where EW$_{DIB}$ is the EW of the particular DIB line in the Milky Way
at log $N({\blHI})=$20.3, the second factor is due to the slope of the
known Milky Way relation between the log EW of the {$\lambda$}5780 DIB
line and the log $N({\blHI})$ of the cloud, and a linear relationship
with metallicity is assumed.  A [Zn/H] of $-1$ is applied when
metallicity is not known (standard DLA metallicity).  As done with the
``Milky Way'' predicted DIB strengths, all of the DIB lines are scaled
by their relative strengths in the Milky Way, and the {$\lambda$}4428
DIB line is further scaled by a factor of 25{\%}.  These results are
noted as ``Z$-$Scaled'' in Table 1.  The right panel of Figure 1 shows
our predictions of the expected {$\lambda$}5780 DIB strength using Eq
2.1.  The data all lie above the 1:1 correlation line.  To find if
metallicity is responsible for the DIB deficiency we require
additional data to adequately constrain our limits.

There are several potential scenarios that can inhibit DIB strength in
DLAs.  If DIB strength scales with metallicity, we would expect to
detect them with higher S/N data.  Perhaps the ionizing radiation
field in regions probed by QSO sightlines destroy these molecules.
Another possibility is that the covering factor of DIB absorbing gas
is much smaller than the covering factor in DLA regions in which case
the QSO ``pencil-beam'' technique is truly hit or miss.  Also, it may
be that our DLAs are not dusty enough to contain DIBs.  In conclusion,
placing DIB absorbers on a substantial statistical footing may be a
difficult goal to realize, and using DIB strengths as a proxy for
$N({\blHI})$ to find intermediate redshift DLAs does not hold much
promise.

\begin{acknowledgments}
B.L. would like to acknowledge a small grant from the IAU and support
from NASA's GSRP.  We would like to thank Michael Murphy, Wal Sargent,
Michael Rauch for contributing the UVES and HIRES spectra to the new
work presented herein.  We would also like to thank Chris Benn for
assisting with the WHT data.
\end{acknowledgments}

\end{document}